\documentclass[usenatbib]{basi}
\usepackage[T1]{fontenc}
\usepackage[british]{babel}
\usepackage[varg]{txfonts}
\usepackage{rotating}
\usepackage{dcolumn}
\begin{document}
\title[Gaia-ESO and multiple analysis pipelines]{Stellar atmospheric parameters and the \emph{Gaia}-ESO Survey experience with multiple analysis pipelines\thanks{Based on data products from observations made with ESO Telescopes at the La Silla Paranal Observatory under programme ID 188.B-3002.}}
\author[R.~Smiljanic, A.~J.~Korn, A.~R.~Casey]%
       {R.~Smiljanic$^1$\thanks{email: \texttt{rsmiljanic@camk.edu.pl}},
       A.~J.~Korn$^2$, A.~R.~Casey$^{3}$, and the \emph{Gaia}-ESO Survey consortium$^{4}$\\
       $^1$Nicolaus Copernicus Astronomical Center, Polish Academy of Sciences, Bartycka 18, 00-716, Warsaw, Poland\\
       $^2$Department of Physics and Astronomy, Uppsala University, Box 516, SE-751 20 Uppsala, Sweden\\
       $^3$School of Physics and Astronomy, Monash University, Australia\\
       $^4$ http://www.gaia-eso.eu}

\pubyear{2017}
\volume{00}
\pagerange{\pageref{firstpage}--\pageref{lastpage}}

\date{Received --- ; accepted ---}

\maketitle
\label{firstpage}

\begin{abstract}
To use libraries of observed stellar spectra, one needs to know the atmospheric parameters of the stars associated to those spectra. It is, however, hard to know what are the real levels of precision and accuracy of these parameters. To overcome this very same problem in a stellar survey, the \emph{Gaia}-ESO Survey implements a comprehensive calibration effort together with the use of multiple parallel analysis methodologies. Here, we discuss the multiple analyses approach and the effort to understand the accuracy and precision of the provided stellar atmospheric parameters.
\end{abstract}


\section{The \emph{Gaia}-ESO Survey}\label{sec:ges}

The \emph{Gaia}-ESO Survey \citep{2012Msngr.147...25G,2013Msngr.154...47R} is a public stellar spectroscopic survey using FLAMES \citep[Fibre Large Array Multi Element Spectrograph;][]{2002Msngr.110....1P}, at the Very Large Telescope (VLT) of the European Southern Observatory (ESO), to obtain medium- and high-resolution spectra of more than 100\,000 stars. \emph{Gaia}-ESO is the major ongoing European stellar spectroscopic survey aimed at complementing \emph{Gaia} \citep{2016A&A...595A...1G} with ground-based spectroscopy. It is the only stellar spectroscopic survey currently making use of an 8-meter class telescope (and thus able to observe stars fainter than in other similar efforts). As a public survey, all raw data obtained by \emph{Gaia}-ESO is immediately available to anyone through the ESO data archive. Releases of advanced data products (radial velocities, atmospheric parameters, chemical abundances, and more) are also made available through ESO\footnote{Through ESO catalog facility at: http://www.eso.org/qi/}.

\emph{Gaia}-ESO is systematically observing stars in all major Galactic components (halo, bulge, thin and thick disks), in globular clusters, and in a large number of open clusters ($\sim$ 65). The open clusters have been selected to cover the parameter space of age, total stellar mass, Galactocentric distance, and metallicity. The stars targeted by \emph{Gaia}-ESO include early- and late-type stars (from O- to M-type), metal-poor and metal-rich stars, giants, dwarfs, and pre-main-sequence stars. A comprehensive calibration strategy has been developed to ensure the analysis of this variety of targets can deliver consistent results \citep[see][]{2017A&A...598A...5P}. 

\section{Multiple pipelines strategy}\label{sec:pipelines}

Multiple analysis pipelines are needed because of the broad range of parameters covered by the stars observed in \emph{Gaia}-ESO and the use of different spectroscopic setups (different resolution and wavelength coverage). In the \emph{Gaia}-ESO consortium, five Working Groups (WGs) manage the spectrum analysis: (i) WG10 deals with the analysis of medium-resolution spectra of FGK-type stars; (ii) WG11 with high-resolution spectra of FGK-type stars; (iii) WG12 with pre-main-sequence stars; (iv) WG13 with early-type stars; and (v) WG14 with non-standard objects.

The focus of this contribution is on the high-resolution spectra of FGK-type stars (WG11). This sample still contains stars with a broad range of parameters: from metal-poor to metal-rich stars; from dwarfs to red giants; and from relatively cool ($\sim$ 4000K) to relatively warm stars ($\sim$ 7000K). Therefore multiple pipelines are still needed even within this WG \citep{2014A&A...570A.122S}.

No single analysis strategy would perform equally well in all these regions of the parameter space. Results from a single pipeline will have different systematics for different types of stars. With a multiple analysis strategy, we can identify the different pipelines that perform well in a given region of the parameter space \citep[see the extensive discussion in][]{2014A&A...570A.122S} and give preference to those results.

\section{Towards accurate and precise parameters}\label{sec:accuracy}

The multiple pipelines strategy introduces some complexity to the understanding of the results. The challenges include how to decide which pipelines perform well in which regions of the parameter space and how to guarantee that the final results for different types of stars are consistent.

The answers to these challenges lie on the calibrator stars. The \emph{Gaia}-ESO calibrators include:

\begin{itemize}
\item stars for which the fundamental parameters are know a priori, such as the \emph{Gaia} benchmark stars \citep{2015A&A...582A..49H,2014A&A...564A.133J} and stars with asteroseismic constraints in $\log~g$ \citep{2016AN....337..970V}. Such stars can be used to evaluate the accuracy of the results;
\item stars for which similar results should be found (i.e., member stars of clusters that should have the same metallicities and chemical abundances);
\item stars where a given behaviour between $T_{\rm eff}$ and $\log~g$ is expected (i.e., clusters stars following the isochrone for the expected cluster age);
\item and stars observed multiple times with different exposure times (i.e., with different signal-to-noise ratio) or with available archival data from different instruments (allowing repeated analyses that can be used to test the precision of the pipelines).
\end{itemize}

\begin{figure}
\centerline{\includegraphics[width=11cm]{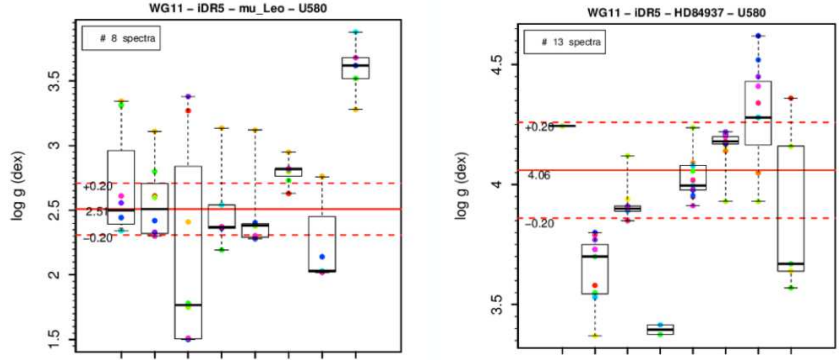}}
\caption{\small{Box plots of the $\log~g$ results obtained from each pipeline (unnamed) in the analysis of 8 spectra of the metal-rich giant $\mu$ Leo (left panel) and 13 spectra of the metal-poor dwarf HD 84937 (right panel). The solid red line is the adopted true value. The precision (size of the boxplot) and accuracy (closeness to the true value) of most pipelines clearly change from one star to another}.\label{f:one}}
\end{figure}

To be useful for calibration, these samples must be analysed in the same way as any other spectra. The changes in accuracy and precision of the multiple pipelines in different regions of the parameter space are exemplified in Fig.\ \ref{f:one}. 

In \emph{Gaia}-ESO, the process of understanding and combining the multiple measurements into a final consistent and homogeneous scale has been called ``homogenisation'', which happens in two stages. The first stage deals with results of multiple pipelines applied to the same sub-dataset (certain stellar type and grating). As currently applied for the high-resolution \emph{Gaia}-ESO spectra, this process aims to infer the real random and systematic errors of each pipeline and to produce an ensemble measurement per star, combining the best aspects of the different analysis techniques and ensuring the accuracy of the results (Casey et al., in prep.). The second stage of homogenisation aims to put together the results obtained for different sub-datasets (for different stellar types and gratings) and ensures they are all on a single final consistent parameter scale. 

The system defined by the atmospheric parameters of $\sim$ 30 benchmark stars is what defines the atmospheric parameter scale of the \emph{Gaia}-ESO results. This is a considerable improvement with respect to the standard approach of using the Sun as the only reference. The \emph{Gaia} benchmark stars are distributed across the parameter space and can provide better references for stars that are not solar like. The full power of the multiple pipeline strategy, and the proper characterisation of the accuracy and precision of the \emph{Gaia}-ESO results, is only obtained thanks to the immense effort put in choosing, observing, and analysing the calibrators. Learning from such effort would be useful for any stellar spectroscopic survey or stellar spectral library that aims to provide accurate and precise atmospheric parameters, in particular for samples with a broad range of parameters.

\section{Summary and take away message}\label{sec:summary}

The \emph{Gaia}-ESO Survey puts a considerable effort into the calibration of its final results. By combining the analysis of calibrators with the multiple analysis strategy, \emph{Gaia}-ESO aims to characterise, and provide, both random and systematic uncertainties in the stellar parameters on a survey scale. This is of utmost importance when results for stars covering a broad parameter space are to be provided in a single consistent scale.

For other stellar surveys or stellar libraries, the \emph{Gaia}-ESO experience offers the following valuable lessons: (i) a comprehensive calibration strategy should be designed; (ii) if possible, it is very useful and informative to apply more than one method of analysis; and (iii) an effort should be made to understand the precision and accuracy of the results.

\section*{Acknowledgements}

RS acknowledges support from NCN (2014/15/B/ST9/03981) and the Polish Ministry of Science and Higher Education. AJK acknowledges support from the Swedish National Space Board.

\label{lastpage}
\end{document}